# Decentralized Micro Water-Energy Co-Optimization for Small Communities

Jesus Silva-Rodriguez and Xingpeng Li
Department of Electrical and Computer Engineering
University of Houston
Houston, TX, USA
{jasilvarodriguez, xli82}@uh.edu

*Abstract*—The water-energy nexus encompasses the interdependencies between water and energy resources identifying the existing links between the production and distribution of these resources. Therefore, understanding the water-energy nexus is crucial for developing sustainable and integrated resource management approaches. This paper proposes a decentralized co-optimization model for a micro water-energy nexus system (MWEN), aiming to optimize the combined supply of both resources to end consumers. The approach respects the separate ownership and management of the water and energy sectors while bridging the gap between their optimized operations. An enhanced version of the alternating direction method of multipliers (ADMM) is proposed, the objective-based ADMM (OB-ADMM), which is able to robustly optimize each system independently towards a common objective, only sharing information about the power consumption of water management, providing privacy for each resource provider.

*Index Terms*—ADMM, Decentralized Co-Optimization, Objective-Based, Water-Energy Nexus.

## NOMENCLATURE

TABLE I: Sets and Indices

| | |
|---|---|
| Generator units | $g \in G$ |
| Energy storage units | $b \in ES$ |
| Water treatment units | $w \in WT$ |
| Water storage tanks | $k \in ST$ |
| Time intervals | $t \in T$ |

TABLE II: Micro Water Management Variables

| | |
|---|---|
| Wastewater treatment flow rate. | $W_t^{WW}$ |
| Wastewater treatment binary status variable. | $u_t^{WW}$ |
| Wastewater reservoir level | $WL_t^{WW}$ |
| Wastewater treatment power consumption | $P_t^{WW}$ |
| Water flow rate of each treatment unit. | $W_{w,t}^{WT}$ |
| Binary status variable of each treatment unit. | $u_{w,t}^{WT}$ |
| Power consumption of each treatment unit. | $P_t^{WW}$ |
| Storage tank water inflow rate. | $W_{k,t}^{STc}$ |
| Storage tank water outflow rate. | $W_{k,t}^{STd}$ |
| Storage tank inflow binary status variable. | $e_{k,t}^{STc}$ |
| Storage tank outflow binary status variable. | $e_{k,t}^{STd}$ |
| Storage tank water level. | $WL_{k,t}^{ST}$ |

TABLE III: Microgrid Energy Management Variables

| | |
|---|---|
| Power output of each generator | $P_{g,t}^G$ |
| Generator binary status variable. | $u_{g,t}^G$ |
| Power input of each energy storage unit | $P_{b,t}^{ESc}$ |
| Power output of each energy storage unit | $P_{b,t}^{ESd}$ |
| Energy storage unit charging status binary variable. | $e_{b,t}^{ESc}$ |
| Energy storage unit discharging status binary variable. | $e_{b,t}^{ESc}$ |
| Energy storage unit charge level. | $EL_{b,t}^{ES}$ |
| Main grid power import. | $P_t^{grid+}$ |
| Main grid power export. | $P_t^{grid-}$ |
| Main grid import status binary variable. | $p_t^+$ |
| Main grid export status binary variable. | $p_t^-$ |

TABLE IV: Micro Water Management Parameters

| | |
|---|---|
| Wastewater treatment minimum flow rate. | $W_{min}^{WW}$ |
| Wastewater treatment maximum flow rate. | $W_{max}^{WW}$ |
| Reclaimed wastewater rate at each time interval. | $WR_t^{WW}$ |
| Wastewater reservoir capacity limit. | $WL_{lim}^{WW}$ |
| Wastewater treatment energy intensity coefficient. | $C^{WW}$ |
| Minimum flow rate of each treatment unit. | $W_{min,w}^{WT}$ |
| Maximum flow rate of each treatment unit. | $W_{max,w}^{WT}$ |
| Energy intensity coefficient of each treatment unit. | $C_w^{WT}$ |
| Minimum flow rate of each storage tank. | $W_{min,k}^{ST}$ |
| Maximum flow rate of each storage tank. | $W_{max,k}^{ST}$ |
| Storage tank minimum capacity | $WL_{min,k}^{ST}$ |
| Storage tank maximum capacity | $WL_{max,k}^{ST}$ |
| Water demand at each time interval. | $W_t^L$ |

TABLE V: Microgrid Energy Management Parameters

| | |
|---|---|
| No-Load cost of each generator | $NL_g^G$ |
| Operation cost of each generator | $C_g^G$ |
| Generator minimum power output. | $P_{min,g}^G$ |
| Generator maximum power output. | $P_{max,g}^G$ |
| Energy storage unit rated power. | $P_{lim,b}^{ES}$ |
| Energy storage unit charging efficiency. | $\eta_b^{ESc}$ |
| Energy storage unit discharging efficiency. | $\eta_b^{ESd}$ |
| Main grid tie-line limit. | $P_{lim}^{grid}$ |
| Combined renewable generation at each time interval | $P_t^{RES}$ |
| Power demand at each time interval | $P_t^L$ |

## I. INTRODUCTION

Potable water and electrical energy distribution systems have traditional been modeled individually, each optimizing its own processes based on their own separate objectives. However, according to current literature, clean water production depends on electrical energy and electrical energy generation depends on water [1]. In a large scale, the links between electricity and water can involve a variety of different functions and processes [2]. In a micro scale, on the other hand, such as the case of small urban communities, the water consumption for electrical generation purposes can be limited to a few components such as thermal generators and hydrogen energy storage, and some renewable sources may use water at production and maintenance stages as well [3]. However, for day-ahead economic dispatch of water and electricity, these links may be limited to simply the power consumption of water-related components such as the treatment processes and water distribution pumps [4]-[6]. Because of these inherent interdependencies, analysis and co-optimization of the distribution of both resources can yield more efficient and cost-effective solutions for managing potable water and electrical energy. This water-energy interdependencies concept is known as the water-energy nexus (WEN) [1].

The economic dispatch concept has widely been used to model microgrid energy management (MEM) problems, under which different energy sources are controlled to fully supply their energy demand achieving minimal operation costs [7]-[8]. Similarly, the economic dispatch approach has already been considered for the management of multiple water sources [9]. Therefore, water and energy systems can be jointly co-optimized under the same management model, simultaneously managing water and energy sources to meet their respective demands in a collaborative manner, bringing the operation cost of both systems to a collective minimum, achieving benefits for both parties.

A WEN system for small communities considering wastewater treatment and clean water storage systems combined with a MEM of multiple energy sources has already been considered in [4], and [10] expanded the concept more, considering a network of these WEN systems along with more water sources as well as modeling the water pumps power consumption. Moreover, similar co-optimization of micro WEN (MWEN) systems were considered in [5] and [6], however the pumps power consumption was the only link considered between the water and energy resources. Furthermore, the MWEN model was presented in [11] with the water network having the potential to provide demand response services so that power consumption of the micro water management (MWM) can be used as a resource to manage the intermittency of renewable generation.

The main consensus found in the literature pertaining to the MWEN is that there are noticeable economic and reliability benefits in the co-optimization of water and energy distribution [1]-[7], [10]-[11]. However, one significant gap in these WEN studies lies in the fact that these combined systems function as a consolidated water-energy system controlled under one central management. Therefore, the idea of a WEN is currently limited to this assumption, which differs from the current management an operation practices of water and energy systems since these are largely separate industries operated by different entities. As a result, this paper proposes a decentralized MWEN for small communities in which the water and energy operators can participate in the nexus to achieve these economic and reliability benefits while still retaining their independence and privacy from one another. The decentralized MWEN model is based directly on the centralized formulation of the MWEN, and is derived by adapting the alternating direction method of multipliers (ADMM) [12], which solves the MWEN problem iteratively. Additionally, an objective-based approach to the ADMM is proposed and applied to the MWEN problem to achieve guaranteed convergence to a feasible solution with a high degree of optimality.

The rest of the paper is organized as follows. Section II presents the formulation of the separate MEM and MWM optimization models as well as the consolidated formulation of the centralized MWEN as a mixed-integer linear program (MILP) achieved by the linearization of the pumps power consumption function by means of a heuristic's method for piecewise linearization [13]. Section III presents the formulation of the decentralized MWEN using ADMM, and the modification of the algorithm into the objective-based ADMM (OB-ADMM) to increase convergence robustness. Section IV presents some case studies that are used to test the OB-ADMM algorithm compared to the standard ADMM, as well as to compare the results of this decentralized approach to the benchmark centralized MWEN. Lastly, Section V concludes the paper.

## II. MICRO WATER-ENERGY NEXUS MODEL

A MWEN system follows the same modeling strategy of economic dispatch used for the modeling of MEM and MWM systems [7], [9]. Moreover, as a combination of the energy and water management systems, the resulting water-energy co-optimization model combines the model constraints of each management system under a single consolidated objective, forming the centralized MWEN co-optimization model.

### A. Microgrid Energy Management

The MEM system is modeled with the objective of minimizing its total operation cost, given by (1), where $\Delta t$ represents the time step between each time interval. In addition, the MEM operation is subject to its system and component constraints.

Generator power output is regulated by (2). Charging and discharging rates of each energy storage unit are regulated by (3) and (4), respectively, charging and discharging status is controlled with (5), while (6) and (7) regulate the charge level of each unit. The tie-line limits with the main grid are enforced with (8) and (9), and (10) controls the import/export status. The power balance constraint is modeled by (11), and it involves power demand, renewable generation, and water management power consumption as part of its "net load," defined by (12).

$$\min C_E = \sum_{t \in T} \Delta t \cdot \left\{ \sum_{g \in G} \left( NL_g^G u_{g,t}^G + C_g^G P_{g,t}^G \right) + C_t^{grid+} P_t^{grid+} - C_t^{grid-} P_t^{grid-} \right\} \quad (1)$$

$$P_{min,t}^G u_{g,t}^G \leq P_{g,t}^G \leq P_{max,t}^G u_{g,t}^G, \forall g \in G, t \in T \quad (2)$$

$$0 \leq P_{b,t}^{ESc} \leq P_{lim,b}^{ES} e_{b,t}^{ESc}, \forall b \in ES, t \in T \quad (3)$$

$$0 \leq P_{b,t}^{ESd} \leq P_{lim,b}^{ES} e_{b,t}^{ESd}, \forall b \in ES, t \in T \quad (4)$$

$$e_{b,t}^{ESc} + e_{b,t}^{ESd} \leq 1, \forall b \in ES, t \in T \quad (5)$$

$$EL_{b,t}^{ES} = EL_{b,t-1}^{ES} + \Delta t \cdot (\eta_b^{ESc} P_{b,t}^{ESc} - P_{b,t}^{ESd}/\eta_b^{ESd}) , \forall b \in ES, t \in T \quad (6)$$

$$EL_{min,b}^{ES} \leq EL_{b,t}^{ES} \leq EL_{max,b}^{ES} , \forall b \in ES, t \in T \quad (7)$$

$$0 \leq P_t^{grid+} \leq P_{lim}^{grid} p_t^+ , \forall t \in T \quad (8)$$

$$0 \leq P_t^{grid-} \leq P_{lim}^{grid} p_t^- , \forall t \in T \quad (9)$$

$$p_t^+ + p_t^- \leq 1 , \forall t \in T \quad (10)$$

$$\sum_{g \in G} P_{g,t}^G + \sum_{b \in ES}[P_{b,t}^{ESd} - P_{b,t}^{ESc}] + P_t^{grid+} - P_t^{grid-} = P_t^{net} , \forall t \in T \quad (11)$$

$$P_t^{net} = P_t^L - P_t^{RES} + P_t^{water} , \forall t \in T \quad (12)$$

### B. Micro Water Management

The MWM system is modeled with a different objective than the MEM. Since it is assumed that the water system does not have any prior knowledge of the forecasted energy prices, then it is best suited to have the objective to minimize the overall energy consumption of all its components. In this case this consumption corresponds to the combined electrical energy consumed by the treatment processes and water pumps. This objective function is given by (13)-(14).

$$\min f_W = \sum_{t \in T} P_t^{water} \quad (13)$$

$$P_t^{water} = P_t^{WW} + \sum_{w \in WT} P_{w,t}^{WT} + P_{pump,t}^{WW} + \sum_{w \in WT} P_{pump,w,t}^{WT} + \sum_{k \in ST} P_{pump,k,t}^{ST} , \forall t \in T \quad (14)$$

The system constraints of the MWM correspond to the operational limits of its treatment units, water storage, and pumps. It is assumed that wastewater treatment is a special case in which the "source" of water is limited, since it depends on the amount of reclaimed water via sewage. Therefore, its constraints are given separate, and these are its flow rate output limits, given by (15), its untreated water reservoir level limits, given by (16)-(17), and its power consumption, given by (18). Any other treatment units, such as groundwater treatment and water desalination, are given by (19) and (20), which define the flow rate output limits and power consumption, respectively. Additionally, the water storage system also present flow rate limits, corresponding to their inflow and outflow rates, regulated by (21) and (22), respectively; its flow status is regulated by (23), and its clean water reservoir limits are regulated by (24)-(25). Lastly, the "water balance," the constraint that ensures enough water is being dispatched to meet demand, is defined by (26).

$$W_{min}^{WW} u_t^{WW} \leq W_t^{WW} \leq W_{max}^{WW} u_t^{WW} , \forall t \in T \quad (15)$$

$$WL_t^{WW} = WL_{t-1}^{WW} + \Delta t \cdot (WR_t^{WW} - W_t^{WW}) , \forall t \in T \quad (16)$$

$$0 \leq WL_t^{WW} \leq WL_{lim}^{WW} , \forall t \in T \quad (17)$$

$$P_t^{WW} = C^{WW} W_t^{WW} , \forall t \in T \quad (18)$$

$$W_{min,w}^{WT} u_{w,t}^{WT} \leq W_{w,t}^{WT} \leq W_{max,w}^{WT} u_{w,t}^{WT} , \forall w \in WT, t \in T \quad (19)$$

$$P_{w,t}^{WT} = C_w^{WT} W_{w,t}^{WT} , \forall w \in WT, t \in T \quad (20)$$

$$W_{min,k}^{ST} s_{k,t}^{STc} \leq W_{k,t}^{STc} \leq W_{max,k}^{ST} s_{k,t}^{STc} , \forall k \in ST, t \in T \quad (21)$$

$$0 \leq W_{k,t}^{STd} \leq W_{max,k}^{ST} s_{k,t}^{STd} , \forall k \in ST, t \in T \quad (22)$$

$$s_{k,t}^{STc} + s_{k,t}^{STd} \leq 1 , \forall k \in ST, t \in T \quad (23)$$

$$WL_{k,t}^{ST} = WL_{k,t-1}^{ST} + \Delta t \cdot (W_{k,t}^{STc} - W_{k,t}^{STd}) , \forall k \in ST, t \in T \quad (24)$$

$$WL_{min,k}^{ST} \leq WL_{k,t}^{ST} \leq WL_{max,k}^{ST} , \forall k \in ST, t \in T \quad (25)$$

$$W_t^{WW} + \sum_{w \in WT} W_{w,t}^{WT} + \sum_{k \in ST}[W_{k,t}^{STd} - W_{k,t}^{STc}] = W_t^L , \forall t \in T \quad (26)$$

The power consumption of pumps is based on their performance characteristics. Pumps operate based on their "pump curve," which is a relation between hydraulic head gain (pressure) and their output flow rate [14]. This pump curve is normally given by pump manufacturers, and it is obtained experimentally. For pumping clean water, the electrical power consumed can be described by (27), which is a convex quadratic function of flow rate where $c_1$, $c_2$ and $c_3$ are constant coefficients related to the pump characteristics.

$$P_{pump} = c_1 W^2 + c_2 W + c_3 \quad (27)$$

Even though (27) is a convex function, it will not be convex when implemented as an equality constraint since quadratic equality constraints are non-affine functions [15]. Therefore, this constraint must be linearized to be able to use ADMM for the combined water-energy management formulation. This linearization is carried out implementing a heuristics least-squares method that formulates a quadratic optimization problem that fits multiple linear functions to input data, creating a piecewise linear fit [13]. Basically, (27) will now be replaced by the following set, implemented for different ranges of the variable $W$:

$$F = \{aW + b\}^{\hat{v}} \quad (28)$$

where $\hat{v}$ is the number of linear functions in the piecewise set $F$.

The linearization program that provides the constraints needed to implement the piecewise linear fit and substitute (27) is shown next [16], where $m$ is the number of data points from the pump performance characteristics dataset, $W \in \mathbf{R}^m$ and $Y \in \mathbf{R}^m$ are the water flow rate and power consumption data points, respectively, $P_{pump} \in \mathbf{R}^{m*\hat{v}-1}$ is the approximated pump power consumption, $a, b \in \mathbf{R}^{\hat{v}}$ are the coefficients of each linear function in $F$, $\alpha \in \mathbf{R}^{m*\hat{v}-1}$ is a binary activation variable to determine what function will be used for a particular value of $W$, and $\Omega$ is a large number.

$$\min_{a,b} \quad \sum_{i=1}^{m}(P_{pump,\hat{v}-1}^i - Y_i)^2$$

s.t.
- $a_1 W_i + b_1 \leq P_{pump,1}^i \leq a_1 W_i + b_1 + \alpha_1^i \Omega$, $\forall i = 1, \dots, m$
- $a_2 W_i + b_2 \leq P_{pump,1}^i \leq a_2 W_i + b_2 + (1 - \alpha_1^i)\Omega$, $\forall i = 1, \dots, m$
- $P_{pump,v-2}^i \leq P_{pump,v-1}^i \leq P_{pump,v-2}^i + \alpha_{v-1}^i \Omega$, $\forall i = 1, \dots, m, v = 3, \dots, \hat{v}$
- $a_v W^i + b_v \leq P_{pump,v-1}^i \leq a_v W^i + b_v + (1 - \alpha_{v-1}^i)\Omega$, $\forall i = 1, \dots, m, v = 3, \dots, \hat{v}$

Solving this optimization problem will provide the coefficients $a$ and $b$ for each of these constraints that can then be used to directly replace the quadratic equality constraints of the pumps power consumption.

### C. Centralized Micro Water-Energy Management

Combining the management operation of the MWM and MEM creates a synergetic system capable of cooperatively optimizing the potable water and electrical power supply, considering all system parameters and constraints under a single optimization problem for a more strategic management of water and energy resources.

The combination of these systems results in a micro water-energy nexus, whose objective is to minimize the operation costs of supplying electric power to the local demand, Therefore, the objective function for the MWEN consists of (1), and implements all the constraints (2)-(12) and (15)-(26), as well as the piecewise linear constraints for the pumps power consumption, with the power consumption of the water management components now included as part of the "net load" in (12). This combined optimization allows for direct coordination of the water management processes so that its power demand can be shifted around during the time period $T$, allowing the MEM to best manage its resources to minimize the total cost, similar to a demand response operation [11].

## III. Decentralized Model via ADMM

### A. ADMM Algorithm Formulation

The alternating direction method of multipliers is a simple but powerful algorithm that is well suited for distributed convex optimization. It takes the form of a decomposition-coordination procedure, in which the solutions to small local subproblems are coordinated to find a solution to a large global problem [12]. For the MWEN, the systems to be separated are the electric power and potable water management systems.

In order to formulate the distributed algorithm via ADMM, it is necessary to identify what constraints involve variables from both participants and label them as "global constraints." These constraints must be relaxed and added to the global objective function of the MWEN in the form of an augmented Lagrangian [12]. These constraints will also involve the global variables that would represent the only piece of information that will be exchanged between the two systems.

The MWEN model is slightly reformulated by duplicating the $P_t^{water}$ into a variable representing the MWM power consumption as predicted by the MEM ($P_{E,t}^{water}$), and a variable representing the actual MWM power consumption determined by the MWM itself ($P_{W,t}^{water}$). Using these duplicated variables, the following constraint is added to the MWEN problem:
$$P_{E,t}^{water} = P_{W,t}^{water}. \quad (29)$$

Equation (29) now represents the global variable, and thus, is used to formulate the augmented Lagrangian, which takes the following form:
$$L = C_E + \sum_{t \in T} \lambda_t \left(P_{E,t}^{water} - P_{W,t}^{water}\right) + \frac{\rho}{2} \sum_{t \in T} \left(P_{E,t}^{water} - P_{W,t}^{water}\right)^2, \quad (30)$$
where, $\lambda$ is the Lagrange multiplier, and $\rho$ is a penalty parameter. This formulation only involves the objective function of the MEM because it is the only function in terms of operation cost. The MWM objective function minimizes power consumption, which in the MWEN is factored into the net load of the MEM getting indirectly minimized to reduce the power generation costs.

The ADMM is executed as an iterative method, optimizing each subsystem in a sequential manner until convergence is reached. In every iteration, the Lagrange multiplier $\lambda$ is updated as shown in (31), and the primal and dual residuals are calculated with (32) and (33) to determine whether the algorithm has converged to an actual solution that respects the global constraint given by (29) [12].
$$\lambda_t^{k+1} = \lambda_t^k + \rho \left(P_{E,t}^{water^{k+1}} - P_{W,t}^{water^{k+1}}\right) \quad (31)$$

$$r_t^k = P_{E,t}^{water^k} - P_{W,t}^{water^k} \quad (32)$$
$$s_t^k = \left(P_{E,t}^{water^k} - P_{W,t}^{water^k}\right) - \left(P_{E,t}^{water^{k-1}} - P_{W,t}^{water^{k-1}}\right) \quad (33)$$

Convergence is achieved ideally when the primal and dual residuals have decreased down to zero. The following metric is defined to measure solution feasibility and be used to determine a stopping point for the ADMM algorithm [18]:
$$\varepsilon^k = \sqrt{\|r^k\|_2^2 + \|s^k\|_2^2}. \quad (34)$$

A standard ADMM algorithm will determine that it has reached convergence when this metric $\varepsilon$ has decreased down to a specified threshold $\varepsilon_{th}$. However, there are some complications that arise depending on the selection of the penalty $\rho$.

The convergence behavior of the ADMM algorithm indicates that for larger values of $\rho$, $\varepsilon$ converges faster. However, as defined by (32)-(34), the solution feasibility metric $\varepsilon$ only guarantees that the solution will be feasible, meaning that (29) is met, but the solution is not guaranteed to be optimal [19]. In fact, as it will be shown in the Case Studies, this particular result occurs for the MWEN for some penalty values. Therefore, additional criteria must be added to the ADMM algorithm to guarantee both feasibility and optimality.

### B. Objective-Based Approach

Besides consideration of the solution feasibility metric $\varepsilon$, the solution optimality must be inspected as well. Therefore, this paper uses the OB-ADMM algorithm previously used in [19]. This objective-based approach follows the following strategy: Once the average objective value's rate of change for a past $k_s$ number of iterations has decreased down to a preset threshold $\beta$; then if at this point the current $\varepsilon$ is below the average $\varepsilon$ for the past $k_s$ iterations, the algorithm stops, and the solution is collected. In this way, the solution is ensured to have a $\varepsilon$ as minimum as possible once the algorithm is within the neighborhood of the optimal solution, ensuring optimality as well as feasibility [19]. The algorithm for the decentralized MWEN via OB-ADMM is represented graphically by Fig. 1.

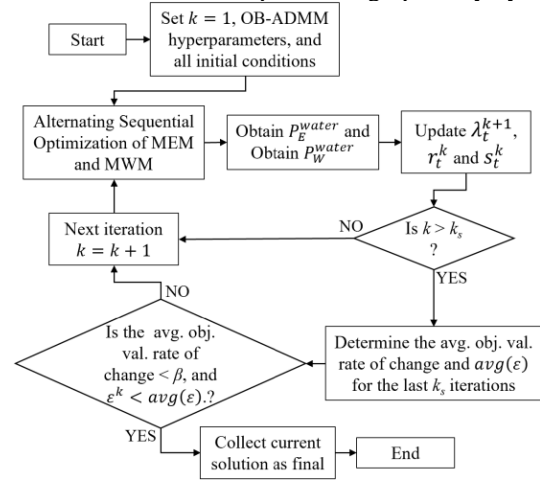

Fig. 1: Flowchart of OB-ADMM algorithm for decentralized MWEN.

## IV. Case Studies

The following test cases are used to 1) test the OB-ADMM algorithm performance and determine appropriate hyperparameters, and 2) compare the results of the decentralized MWEN model via OB-ADMM against the centralized model. All test cases use the same main grid import/export prices based on the average locational marginal prices from ERCOT's datasets [20]. The export prices correspond to 80% of the import prices at all time intervals. Moreover, Tables VI and VII provide parameters for the MEM and MWM components, respectively, that are used in each of the three test cases considered.

TABLE VI: MEM component parameters [24]-[28].

| Source Type | Natural Gas Gen. | Diesel Gen. | BESS |
|---|---|---|---|
| NL Cost [$/h] | 6.00 | 5.00 | - |
| Op. Cost [$/kWh] | 0.04 | 0.24 | - |
| Min. Power [kW] | 40 | 50 | 0 |
| Max. Power [kW] | 400 | 500 | 550 |
| Storage Capacity [kW] | - | - | 2750 |

TABLE VII: MWM component parameters [29]-[32]

| Source Type | Wastewater | Groundwater | Desalination | STU |
|---|---|---|---|---|
| Energy Intensity [kWh/m$^3$] | 52 | 0.154 | 3.6 | - |
| Min. Flow Rate [m$^3$/h] | 0.681 | 0.681 | 0.681 | 0.681 |
| Max. Flow Rate [m$^3$/h] | 2.725 | 2.725 | 2.725 | 2.725 |
| Reservoir Capacity [m$^3$] | 37.854 | - | - | 27.255 |

The first test case consists of 70 residential and 3 commercial units in a suburban area with only one natural gas generator, and a tie-line thermal limit of 1200 kW. Additionally, it features a groundwater treatment plant and one storage tank unit. The second test case consists of 100 residential and 4 commercial units in a coastal area with a diesel generator, and a tie-line thermal limit of 1400 kW. Additionally, it features a water desalination plant and one storage tank unit. The third test case consists of 6 residential and 2 commercial units in a remote, isolated coastal area with no access to the main power grid. This case has a natural gas as well as a diesel generator, and a water desalination plant. All test cases feature a wastewater plant, as well as one battery energy storage unit (BESS) with a round up efficiency of 88.3% and one storage tank unit (STU). Moreover, the power and water demand profiles for each test case are provided in Figs. 2 and 3, respectively, according to the average power and water consumptions of residential and commercial units in Texas [21]-[23].

### A. Algorithm Performance

Test case 1 is taken as reference to analyze the performance of the ADMM for decentralizing the MWEN. Fig. 4 shows the convergence of the solution feasibility and the normalized objective value for a penalty value of $\rho = 0.01$, while TABLE VIII shows the final objective value and solution feasibility results obtained with the standard ADMM using a feasibility threshold $\varepsilon_{th}$ of $10^{-6}$, and TABLE IX shows the results obtained with OB-ADMM using a $\beta = 10^{-4}$ for different $k_s$ values.

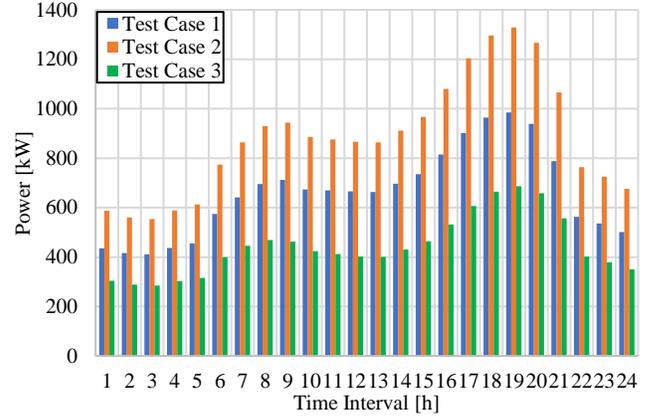

Fig. 2: Test cases power demand profiles.

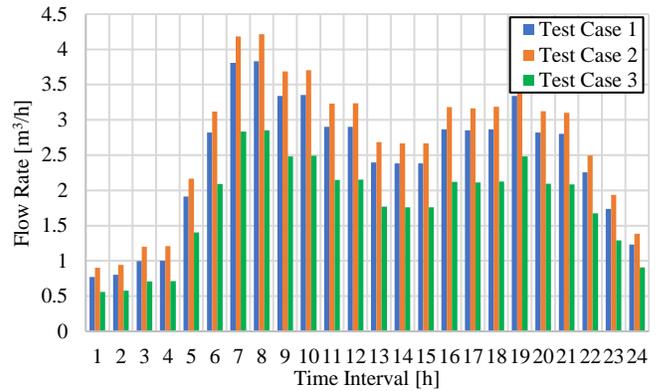

Fig. 3: Test cases water demand profiles.

In these results, the solution feasibility $\varepsilon$ is observed to converge to zero faster than the normalized objective value converges to the optimal. However, comparing the obtained global objective value to that of the centralized MWEN reveals that even though $\varepsilon$ has converged, the stopping point of the standard ADMM has not reached the optimal value, and the deviation is higher with higher values of $\rho$. Therefore, even though solution feasibility is improved with higher values of $\rho$, the solution optimality is affected. This means that simply using $\varepsilon$ as the only convergence criterion is not sufficient, and thus, the objective value must be considered as well to ensure optimality in addition to feasibility.

The OB-ADMM introduces analysis of the objective value to determine convergence to the optimal solution. However, since in practical cases the actual solution of the centralized model is not known, the OB-ADMM checks the rate of change of the objective value since this rate is shown to decrease as the algorithm approaches the neighborhood of the optimal solution, as seen in Fig. 4. Using this approach, the algorithm progresses for a few more iterations until it finds a point that not only presents a low $\varepsilon$, but also an objective value closer to the optimal solution.

The results also indicate that for the OB-ADMM strategy, a penalty of $\rho = 0.01$ seems to be an acceptable choice, and the trend also demonstrates that further increasing $\rho$ negatively affects optimality. Increasing the value of $k_s$ improves the solution, but this comes at the cost of more iterations.

However, the OB-ADMM algorithm is able to reach higher optimality solutions than the standard ADMM by decreasing the objective value by at least a factor of 100. Therefore, OB-ADMM increases algorithm robustness only at the expense of longer computation time.

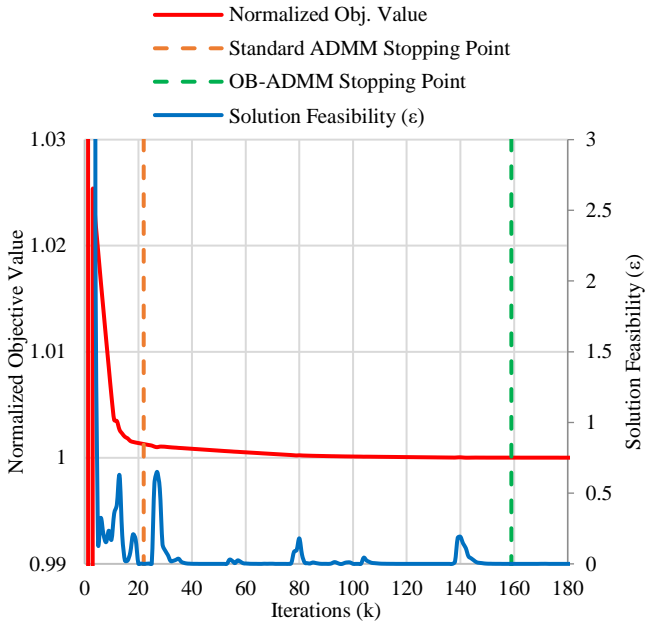

Fig. 4: Normalized objective value and solution feasibility for the MWEN via standard and objective-based ADMM methods.

TABLE VIII: Results for MWEN with Standard ADMM

| Penalty $\rho$ | % Difference w/ Centralized | Solution Feasibility $\varepsilon$ |
|---|---|---|
| 0.01 | 0.13% | $1.60 \times 10^{-13}$ |
| 0.1 | 2.63% | $6.70 \times 10^{-13}$ |
| 1 | 3.02% | $1.31 \times 10^{-13}$ |

TABLE IX: Results for MWEN with OB-ADMM.

| Penalty $\rho$ | Iteration Offset $k_s$ | % Difference w/ Centralized | Solution Feasibility $\varepsilon$ |
|---|---|---|---|
| 0.01 | 10 | 0.00177% | $1.51 \times 10^{-5}$ |
|  | 50 | 0.00121% | $3.66 \times 10^{-13}$ |
|  | 100 | 0.000696% | $6.07 \times 10^{-14}$ |
| 0.1 | 10 | 0.00218% | $5.63 \times 10^{-5}$ |
|  | 50 | 0.00212% | $2.15 \times 10^{-11}$ |
|  | 100 | 0.00209% | $1.81 \times 10^{-13}$ |
| 1 | 10 | 0.101% | $3.01 \times 10^{-5}$ |
|  | 50 | 0.101% | $9.08 \times 10^{-5}$ |
|  | 100 | 0.101% | $8.12 \times 10^{-5}$ |

### B. Test Cases Results

The OB-ADMM algorithm for the decentralized MWEN is used to solve all three test cases using the hyperparameters $\rho = 0.01$, $\beta = 10^{-4}$, and $k_s = 50$. Similarly, the standard ADMM is also used for comparison, with a $\varepsilon_{th} = 10^{-6}$. The objective value results for both algorithms are shown in Fig. 5 compared to the centralized optimal operation costs of $506.01, $885.66, and $1338.13 for test cases 1, 2, and 3, respectively. Additionally, TABLE X presents the results of the total MWM energy consumption in the 24-hour period for every test case, which corresponds to the global variable.

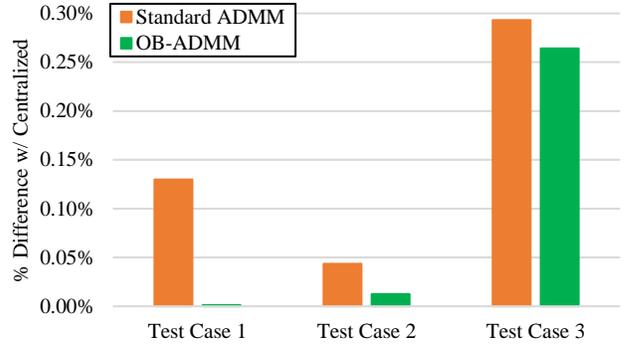

Fig. 5: Objective value deviation from the centralized model for each ADMM approach.

TABLE X: MWM Energy consumption for every test case.

| System | Test Case 1 | Test Case 2 | Test Case 3 |
|---|---|---|---|
| Centralized MWEN | 1813.73 kWh | 1884.58 kWh | 1239.32 kWh |
| Decentralized MWEN (Standard ADMM) | 1808.80 kWh | 1884.35 kWh | 1239.08 kWh |
| % Difference (Standard ADMM) | -0.27% | -0.012% | -0.019% |
| Decentralized MWEN (OB-ADMM) | 1813.13 kWh | 1885.05 kWh | 1238.35 kWh |
| % Difference (OB-ADMM) | -0.033% | 0.025% | -0.078% |

These results show how the OB-ADMM achieves greater solution optimality for all three test cases. In terms of the MWM system energy consumption, the results of test case 1 and 2 show that the standard ADMM achieved lower consumption but at the expense of a higher operation cost. This occurs because there are times when power from the grid is substantially cheaper, and thus it is more convenient to consume power for water processes at those times although at a reduced energy efficiency, which is the solution OB-ADMM obtains. The standard ADMM on the other hand, delivers a less cost-effective solution as a result of its larger deviation from the optimal solution, while the OB-ADMM reaches a point closer to the optimal. However, in the third test case where there is no main grid connection, the OB-ADMM solution is more energy efficient as well as cost-effective.

## V. CONCLUSIONS

This paper proposes a decentralized approach to the micro water-energy nexus system that allows water and energy distribution operators to obtain the same economic, reliability and efficiency benefits of a MWEN while preserving their privacy and separate operations. The paper proposes an enhanced objective-based ADMM approach for distributed optimization which guarantees both feasibility and optimality. A centralized MWEN model is used as the benchmark, and the results of the decentralized MWEN via OB-ADMM are used to corroborate that the proposed model can obtain the same solution without the need for both water and energy systems to be consolidated into a single centralized operator.

The results indicate that both decentralized approaches considered can effectively decentralize the MWEN problem with ADMM, yielding solutions that are very close to the actual optimal solution of the base centralized model. However, the OB-ADMM is able to guarantee a higher degree of optimality for the same selection of penalty values compared to the standard ADMM, only taking additional iterations to ensure the difference in global objective value is minimal, providing the decentralized algorithm with additional convergence robustness compared to that of the standard ADMM approach.